# The role of the Look Elsewhere Effect in determining the significance of an oscillation disappearance search for a light sterile neutrino

Gioacchino Ranucci

*Istituto Nazionale di Fisica Nucleare*

*Via Celoria 16 - 20133 Milano*

*Italy*

Phone: +39-02-50317362

Fax: +39-02-50317617

e-mail: gioacchino.ranucci@mi.infn.it

In the ongoing vibrant experimental quest to assess whether the numerous indications for a light sterile neutrino are only experimental fluctuations or the manifestations of a profound and real underlying effect, one aspect which has recently attracted a specific interest is the statistical treatment of the data. Especially in cases of supposed positive hints, the correct statistical assessment of their significance is of paramount importance, to avoid that potential overstatements lead to a wrong understanding of the real status of the experimental investigation in the field. In this work I show how latest crucial advancements in the statistical data processing for the interpretation of the output of a sterile search can be effectively put and understood in the context of the Look Elsewhere Effect phenomenon, developed and now of routine usage for results interpretation in other areas of HEP research.

# 1. Introduction

The hot subject of the light sterile neutrino has triggered throughout the years a scrutiny of all related theoretical and experimental factors in the quest to unveil the mystery surrounding the contradictory indications coming from the plurality of experiments carried out to try to solve (unsuccessfully so far) the dilemma of its existence. A recent, complete survey on the multiple issues of the field can be found in [1].

A feature which has attracted some attention in latest years is the correct assessment of the significance of a positive hint coming from an experiment of the kind of those performed in this area. The issue is of particular interest when the putative hint is in the troublesome ballpark of a few sigma, well below the golden standard of 5 σ threshold for a discovery, but carrying the tantalizing suggestion that maybe something interesting is happening.

A set of illuminating papers [2],[3],[4] and [5] have shed light on the statistical methods in the sterile neutrino hunt, leading to the acknowledgement that the standard approach based on the most famous and celebrated theorem in statistics, the Wilks' theorem [6], is providing a framework plagued by inaccuracies in the reported results. This outcome is essentially stemming from the non-applicability of the validity conditions required by the Wilks' theorem, which affects both the evaluation of the significance of an alleged detection of a signal, as well as the determination of the related confidence/exclusion intervals.

Building upon these previous results, in this note I show how the determination of the significance in the sterile search can be put in a more general framework encompassing methodologies developed to tackle the so-called Look Elsewhere Effect (LEE) [7] in situations featuring the search of an a-priori unknown signal over a wide range of its possible domain of existence.

# 2. A test set-up for exemplificative calculations

While experiments searching for a light sterile neutrino can be both of appearance and disappearance type, the underlying statistical methodologies to interpret the data are similar. The focus of this work is specifically on the implications of the statistical treatment for the sterile investigation in the framework of disappearance searches. A typical configuration of this type can be modeled through the scheme of a disappearance measurement accomplished by deploying a proper neutrino/antineutrino source close to a liquid scintillator detector, an arrangement that shares many similarities with a standard disappearance reactor sterile experiment.

Specifically, I consider a neutrino source, $^{51}$Cr, at short distance of few meters from a spherical scintillator detector, as shown in figure 1. The shell depicted in the figure stemming from the intersection of the active volume with the incoming neutrino flux shows that the count rate in the detector depends geometrically upon the distance L from the source, and therefore the analysis can be carried out by subdividing such a distance in discrete binning. If the source originates a flux function of the energy, such a set-up would automatically give rise to the standard L/E dependence which characterizes most of the experiments in the field.

The scenario here is made simpler by the occurrence that most of the emitted neutrinos are at around 750 keV (two very close line at 746 and 751 keV), and thus the L/E dependence reduces to a simpler L only dependence, leaving however unaltered the implications for the statistical features of the problem. For completeness it should be added that the $^{51}$Cr source generates also a further minor monochromatic flux at about 430 keV (two lines at 426 and 431 keV), which for the purpose of this exercise has been taken into account assuming a total flux at an averaged energy of 0.714 keV.

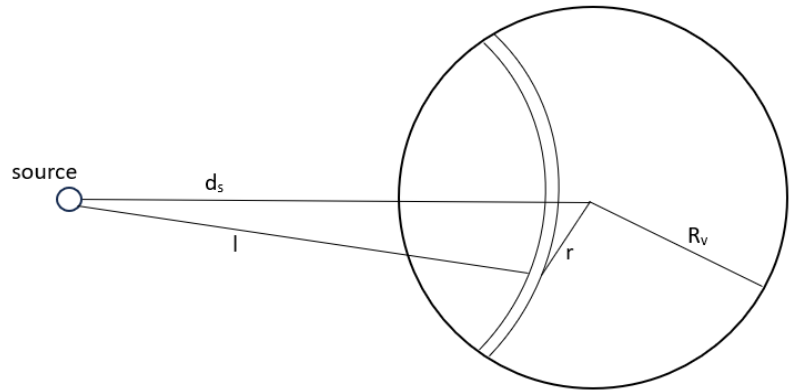

**Figure 1** – *Exemplificative test set-up scheme of a neutrino source deployed at short distance from a spherical liquid scintillator detector, suitable to perform typical statistical calculations representative of a standard disappearance search for a light sterile neutrino.*

The neutrino flux, interacting elastically with the electrons of the scintillator, produces a geometrically varying counting rate while proceedings through the fiducial volume of the active medium, further modulated by the electron neutrino survival probability, in case of existence of the sterile state and the consequent active sterile oscillation, governed by the electron neutrino survival probability $P_{ee}$

$$P_{ee} = 1 - sin^2(2\theta_{ee})sin^2\left(\frac{\Delta m_{41}^2 L}{4E}\right) \quad (1).$$

This is the formulation valid in the simplest so called 3+1 scenario, i.e. three standard neutrinos and a sterile state. The sterile neutrino mixing parameters are the angle $\theta_{ee}$ and the mass squared difference $\Delta m^2_{41}$ [8].

It is very instructive to depict what would be the effect of such a modulation in case of an actual active-sterile oscillation. This is done in figure 2, where the average spatial profile of the count rate is plotted in absence of oscillation, as well as for two cases characterized by $\Delta m^2_{41}=1$ and sen$^2 2\theta_{ee}$=0.05 and 0.9, respectively. The latter, obviously unrealistic since such a large value of the mixing angle is already ruled out, shows an unmistakable signature that would make not relevant any statistical discussion. The former instead demonstrates visually the difficulty to unravel a small effect, which would appear further fuzzy if the Poisson variation of the counts in each bin would be superimposed to the average plot in the figure, explaining the need of an unambiguous criterion to assess in a statistically rigorous way its possible presence.

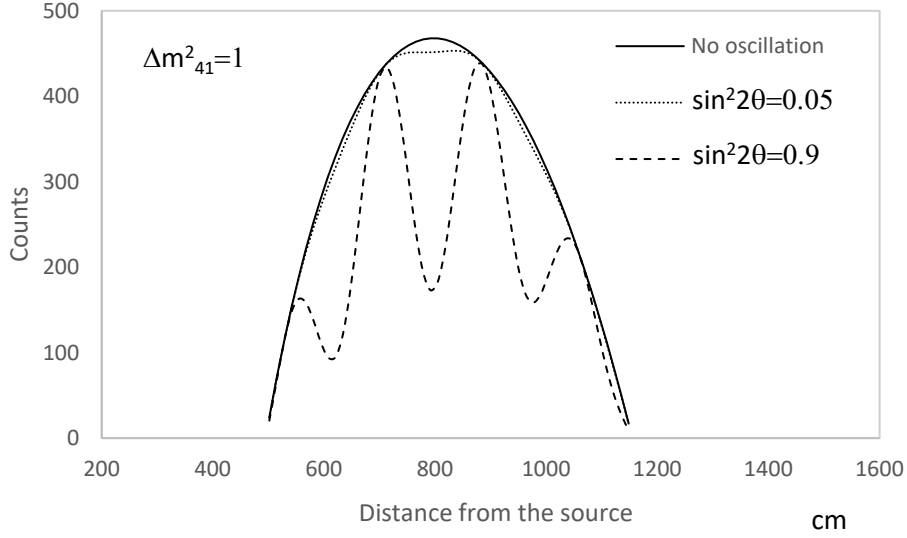

**Figure 2** – *Average counts induced by the source in the example test set-up in three different cases of no oscillation (solid line) and with oscillation (dotted and dashed lines) with parameters indicated in the box.*

It is well known that the detection of an effect (signal) masked by statistical variations in the collected data (noise), when the description of the problem implies the presence of multiple parameters, is in general addressed in statistical theory under the chapter of composite testing problems, through the implementation of a generalized likelihood ratio test (also referred in our field as profile likelihood ratio test) [9][10].

In the case of the investigation for an oscillation effect in the context of a disappearance sterile search, if we want to test the $H_0$ hypothesis of the occurrence of a specific couple of parameters ($\Delta m^2_{41}$, $sen^2 2\theta_{ee}$) (null hypothesis in statistical parlance) against the $H_1$ hypothesis represented by all other possible couples of parameters (alternative hypothesis), the corresponding generalized likelihood ratio is written as

$$GLR = \frac{max_\eta \mathcal{L}(nobs_i; \eta, \Delta m^2_{41} = X, sin^2 \theta_{ee} = Y)}{max_{\eta, \Delta m^2_{41}, \theta_{ee}} \mathcal{L}(nobs_i; \eta, \Delta m^2_{41}, sin^2 2\theta_{ee})} \qquad (2)$$

where $\mathcal{L}$ is the likelihood function of the problem at hand, $nobs_i$ are the observed counts in the generic bin i and η represents a set of nuisance parameters, while X and Y are the specific values of the couple of parameters under test.

In the example under study, I consider a spherical unsegmented scintillator detector whose only important statistical uncertainty is the error on the fiducial volume, realistically taken at the 1% level. Moreover, the source itself features an uncertainty which, with a proper calibration, can be also kept at the 1% level. For simplicity, the error on the fiducial volume is attributed to the source strength, by summing the two individual errors in quadrature, since the effect is the same, i.e. an uncertainty in the total number of detected events. By denoting with $N_0$ the total number of events emitted by the source, which is now the only nuisance parameter, relation (2) becomes

$$GLR = \frac{max_{N_0}\mathcal{L}(nobs_i; N_0, \Delta m_{41}^2 = X, sin^2\theta_{ee} = Y)}{max_{N_0,\Delta m_{41}^2,\theta_{ee}}\mathcal{L}(nobs_i; N_0, \Delta m_{41}^2, sin^2 2\theta_{ee})} \quad (3)$$

In the example the nominal number of emitted neutrinos by the source during the execution of the test is taken in the ballpark of 15000 events, and the spatial segmentation at the level of 50 bins, so that in each bin the sizable number of events implies a reasonable gaussian approximation. Under this approximation, and by taking -2 times the natural log of GLR, equation (3) becomes

$$-2lnGLR = min_{N_o}\left(\sum_i^n\left(\frac{(nobs_i - nmodel_i(N_0, \Delta m_{41}^2 = X, sin^2 2\theta_{ee} = Y))^2}{nobs_i}\right)\right.$$
$$\left.+ \left(\frac{N_0 - N_{0f}}{\sigma_f}\right)^2\right)$$
$$- min_{N_0,\Delta m_{41}^2 sin^2\theta_{ee}}\left(\sum_i^n\left(\frac{(nobs_i - nmodel_i(N_0, \Delta m_{41}^2, sin^2 2\theta_{ee}))^2}{nobs_i}\right)\right.$$
$$\left.+ \left(\frac{N_0 - N_{0f}}{\sigma_f}\right)^2\right) \quad (4)$$

In (4) *nmodel_i* is the prediction of counts in each bin, *N_{0f}* the nominal number of neutrinos emitted by the source and *σ_f* the relevant error, obtained as explained above. Furthermore, the pull factor constraining the number of events from the source is explicitly shown.

Relation (4) is the standard least square formulation stemming from the generalized likelihood ratio test, it is obviously a random variable depending on the observed random variables *nobs_i*, and it is by definition the test statistic used to infer conclusions on the problem under examination.

A general property of the generalized likelihood ratio (2) is that it varies from 1, when the maximum values at denominator and numerator are equal, e.g. the hypothesis (parameters) tested exhibits maximum compatibility with the true one, to 0, signaling instead maximum incompatibility between hypothesis and reality. By taking -2ln of the ratio, maximum compatibility corresponds to 0 and maximum incompatibility to infinity, and this is therefore the range of variability of the test statistic as expressed by the formulation (4). How its values in between are distributed, i.e. their actual PDF, is the matter of the Wilks' theorem, which states that under the null hypothesis, i.e. if $H_0$ is true, -2ln of a generalized likelihood ratio obeys to a Probability Density Function (PDF) coinciding with a $\chi^2$ function with as many degrees of freedom as the difference between the maximization parameters at numerator and denominator, two in the present case, i.e. $sin^2 2\theta_{ee}$ and the mass squared difference $\Delta m^2_{41}$. Such a property, however, is asymptotic, i.e. the sample size should be large enough, and moreover requires some general regularity conditions to hold.

The following discussion encompasses this point, the validity of the Wilks' theorem in the context of the search of an effect a priori unknown and the specific implication of such a scenario in the sterile neutrino quest.

Before that, let's point out that the test statistic (3) or (4), with X and Y different from 0, upon an experimental outcome is used to draw the exclusion/confidence intervals at a certain defined confidence level, and therefore can be named as the "exclusion test statistics", while when the model at numerator of the GLR (3), or equivalently the first term of the least square sum (4) is written without the oscillatory model, then the (3), and hence also (4), becomes the "discovery test statistic", since the H0 hypothesis (null hypothesis) becomes the absence of the signal, and it is the one used to ascertain the significance of a presumptive signal observed in the collected data. The rest of the scrutiny is centered on the properties of the discovery test statistic and how it leads to the manifestation of the Look Elsewhere Effect.

### 3. Discovery test statistic as function of $\Delta m^2_{41}$

In order to unravel the features of the discovery test statistic, now written to explicitly show the absence of the signal in the model in the first term of (4) as

$$min_{N_o}\left(\sum_i^n \left(\frac{(nobs_i - nmodel_i(N_0, no\ signal))^2}{nobs_i}\right) + \left(\frac{N_0 - N_{0f}}{\sigma_f}\right)^2\right)$$
$$- min_{N_0,\Delta m^2_{41} sin^2\theta_{ee}}\left(\sum_i^n \left(\frac{(nobs_i - nmodel_i(N_0, \Delta m^2_{41}, sin^2\theta_{ee}))^2}{nobs_i}\right)\right.$$
$$\left. + \left(\frac{N_0 - N_{0f}}{\sigma_f}\right)^2\right) \qquad (5)$$

let's proceed to simulate the experimental example depicted above in Fig. 1. For this purpose, the length of the active volume is divided into 50 bins. In each simulation, the bin contents are generated assuming the absence of a signal, thus according to the nmodel($N_0$, no signal) function in the first term of the least square members.

Obviously, the minimization of the second term provides the best fit point. It is instructive to illustrate the effect of the minimization as function of $\Delta m^2_{41}$ for a typical outcome of a simulation with no signal. Such a behavior is shown in figure 3, where for each $\Delta m^2_{41}$ the value of the test statistic obtained by minimizing against $sin^2 2\theta_{ee}$ is reported. The peaked feature apparent in Fig. 3 stems from the subtraction of the rapidly varying second term of (5) from the constant first term, and when the second term becomes very low, meaning that a good fit has been achieved, correspondingly a high peak is generated. Therefore, the highest peak in the discovery test statistic plot vs $\Delta m^2_{41}$ is the strongest indication of a putative signal, as it corresponds to the absolute minimum of the second term over the entire ($sen^2 2\theta_{ee}$, $\Delta m^2_{41}$) search region.

In eq. (5) the extra minimization parameters in the second term are two with respect to the first term, $sen^2 2\theta_{ee}$, $\Delta m^2_{41}$, therefore a naïve application of the Wilks' theorem would induce to believe that the PDF of the test statistic (5), i.e. the distribution of the highest peak in Fig. 3, follows a $\chi^2$ function with two degrees of freedom.

In reality, this is not the case, for the following three reasons. First, the Wilks' theorem requires that the parameters kept fixed in the first term assume values well in the interior of their admissible region. Instead, the no signal model at numerator is obtained by imposing the value 0 to sen$^2 2\theta_{ee}$, which therefore lies at the border of its domain, contrary to the requirements of the theorem. Second, the minimization parameters in (5) must all be free, while $N_0$ is instead constrained through the pull term stemming from physical considerations.

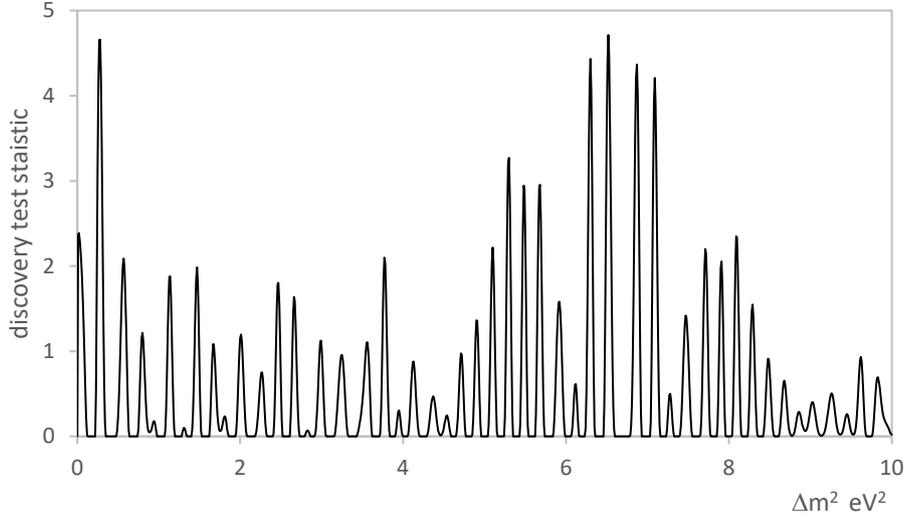

**Figure 3** – *Plot of the test statistic eq. (5) as function of $\Delta m^2_{14}$ after minimization against $sin^2 2\theta_{ee}$.*

Finally, the parameter $\Delta m^2_{41}$ does not have a physical meaning in the background only condition that we want to test; actually, it exists only under the alternative (i.e. at the second term of equation (5)).

Before examining in detail the implications of such violations on the test statistic distribution, it can be highlighted that the mapping of the problem performed as shown in figure 3, amounting to search for the highest peak on a range of $\Delta m^2_{41}$ as indication of the presence of an oscillation effect, leads to a framework fully equivalent to a search of a particle of unknow mass over a wide mass range.

Starting from the seminal paper [11], addressing the issue of hypothesis testing when a parameter exists only under the alternative, such a search particle configuration has been thoroughly studied in [12] and [13], with the main findings being summarized as the demonstration of the non-validity of the Wilks' theorem also in that case and the development of two different, but equivalent methodologies to deal practically with the modification of the distribution of the test statistic from the expected $\chi^2(2)$ distribution, according to an explicative framework which is what goes under the name of Look Elsewhere Effect (LEE). The very name adopted for the effect underlines the occurrence that the search for the signal is performed everywhere in the investigation region, without any a priori knowledge of the mass location; moreover, it naturally leads to the now well-known concepts of local and global significance, the former valid for a particular mass and the latter over the whole mass search interval.

Specifically, the origin of the LEE is traced back to the third of the violations above mentioned, i.e. the existence of a parameter only under the alternative, which leads to a scenario effectively entailing to test individually each value of that parameter in its admissible range, the mass of the particle in [12] and [13], assuming then as test statistics the maximum among all the computed log-likelihood ratios; this effective interpretation is the precise motivation of the name of Look Elsewhere Effect adopted to define the phenomenon under examination.

Now, the above description adheres exactly to the situation depicted in Fig. 3. The study of the present oscillation problem can be, thus, greatly facilitated by resorting to the analogy to the particle of unknown mass search and to the LEE studied in that context.

Before closing this paragraph, it is useful to mention that a recent discussion about the limit of validity of the Wilks' theorem can be found in [14], and other considerations on the LEE in references [15], [16] and [17].

## 4. Local distribution of the test statistic

The first step to extend here the application of the LEE methodology for the search of a new particle is the study of the local distribution of the test statistic obtained for a fixed mass, i.e. for a fixed $\Delta m^2_{41}$ in the oscillation problem under study.

However, to unveil the role played by the minimization parameters it is convenient to divide the MC study into three different steps: a) fully unconstrained, i.e. free $N_0$ and $sen^2 2\theta_{ee}$ not bounded to 0, b) partially constrained, i.e. free $N_0$ and $sen^2 2\theta_{ee}$ bounded to 0, c) fully constrained, i.e. constrained $N_0$ through the pull term and $sen^2 2\theta_{ee}$ bounded to 0.

Several MC tests are thus performed under the three above assumptions of the set-up depicted in Fig. 1, by simulating only noise, i.e. only Poisson variations of the counts around the expected geometrical variations without any oscillation effect.

*4.a Fully unconstrained case*

In the case of unconstrained parameters, the simulation to determine the distribution of the test statistic (5) for fixed $\Delta m^2_{41}$ is performed by removing the pull factor on $N_0$, while the parameter $sen^2 2\theta_{ee}$ is fictitiously allowed to go up and down with respect to 0. Clearly, the latter condition is not physically realistic, but it will help to shed light on the mathematics of the problem.

The results of four simulations performed for the $\Delta m^2_{41}$ values of 0.05, 1, 5 and 10 ev$^2$ are shown in Fig. 4, together with the $\chi^2(1)$ distribution, which is faithfully reproduced in all four cases.

Indeed, the Wilks' theorem conditions are now fulfilled: $\sin^2 2\theta_{ee}=0$ is in the interior of the admissible range, having adopted the trick to remove the physical boundary at 0, and $N_0$ is unconstrained, as well. Moreover, $\Delta m^2_{41}$ does not play any specific role, since it is kept fixed. Therefore, the difference between the minimization parameters in the left and right terms of (5) is only 1, and the perfect reproduction of the $\chi^2(1)$ is the experimental proof of the validity of the Wilks' theorem at any $\Delta m^2_{41}$ value.

It is worth to repeat that this exercise has only the meaning to understand the underlying mathematics leading to the test statistic distribution, since values of $\sin^2 2\theta_{ee}$ less than 0 are obviously unphysical.

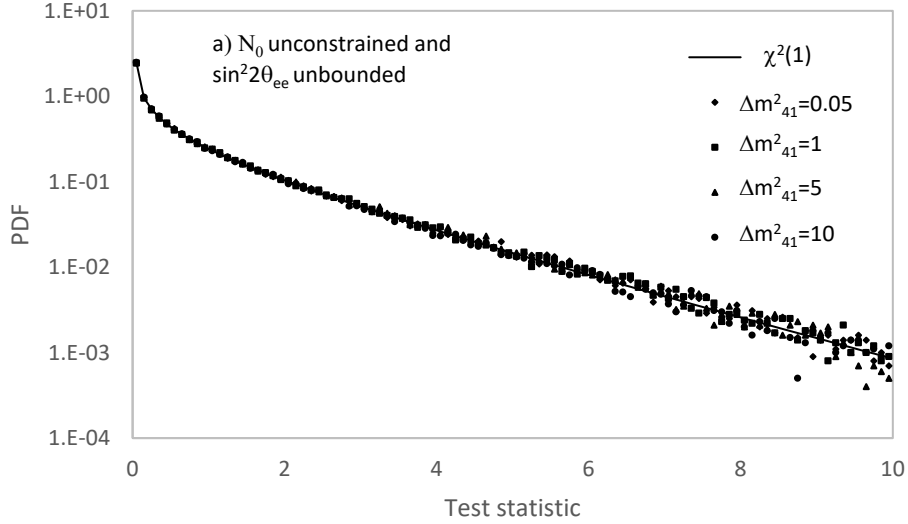

**Figure 4** – *Test statistic MC distributions for four $\Delta m^2_{41}$ values compared with the $\chi^2(1)$ function. The simulations are performed for $N_0$ unconstrained and $sin^2 2\theta_{ee}$ unbounded. The $\chi^2(1)$ function is very well replicated.*

*4.b Partially constrained case*

This configuration is realistic, since $sen^2 2\theta_{ee}$ is restrained only to physically admissible values, while $N_0$ unconstrained entails to ignore any a-priori information about its value; this is not done in practice, but, in any case the resulting situation is in principle allowed.

The violation of the Wilks' theorem is now visible in Fig. 5, showing that the resulting MC distribution for any $\Delta m^2_{41}$ is a ½ $\chi^2(1)$ distribution plus a Dirac delta at 0, accounting for 50% of the outcomes.

Such a result is expected [18] when the amplitude parameter in the fit is bounded to zero and there is no signal in the data; in the present case what happens is that the model, whose amplitude is governed by $sen^2 2\theta_{ee}$, can describe only the downward fluctuations of the counts, since the expected oscillation effect could produce only a depletion of the rate. But, when there is only noise, the fluctuations can be both upward and downward, and thus the model cannot describe the 50% upward occurrences, forcing correspondingly the test statistic to zero. In the other 50% of the occurrences, instead, the same situation as the previous case is replicated, aligning the outcome to a $\chi^2(1)$ distribution. Thus, in summary, the global distribution of the test statistic is a Dirac delta centered at zero, plus a half $\chi^2(1)$ distribution, and this explains the result in Fig. 5.

*4.c Fully constrained case*

The deviation from the expectation of the Wilks' theorem is strongly exacerbated when the constraint is applied to $N_0$, as shown in figure 6, where the results of the simulation are reported in the complete condition, i.e. constraint on $N_0$ and bound on $sen^2 2\theta_{ee}$.

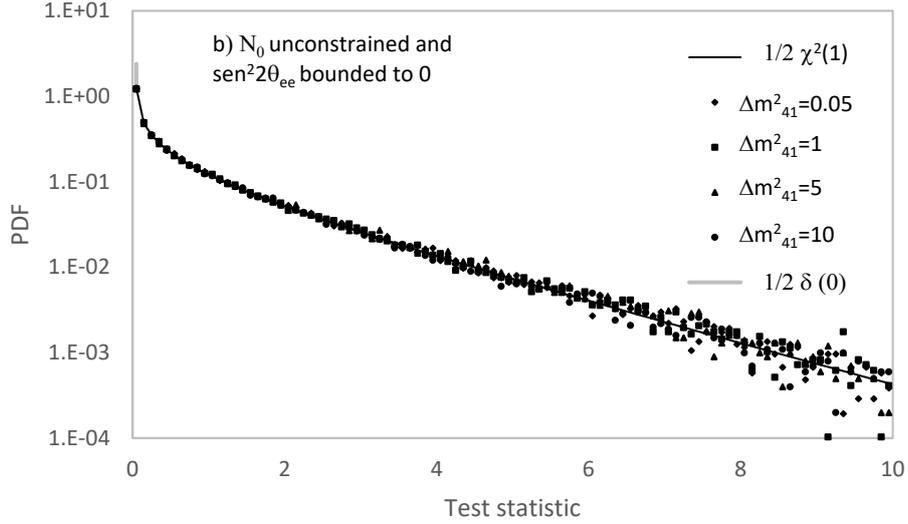

**Figure 5** - *Test statistic MC distributions for four $\Delta m^2_{41}$ values described by a ½ $\delta(0)$ + ½ $\chi^2(1)$ function. The simulations are performed for $N_0$ unconstrained and $\sin^2 2\theta_{ee}$ bounded to 0. The agreement with the reference function is very good.*

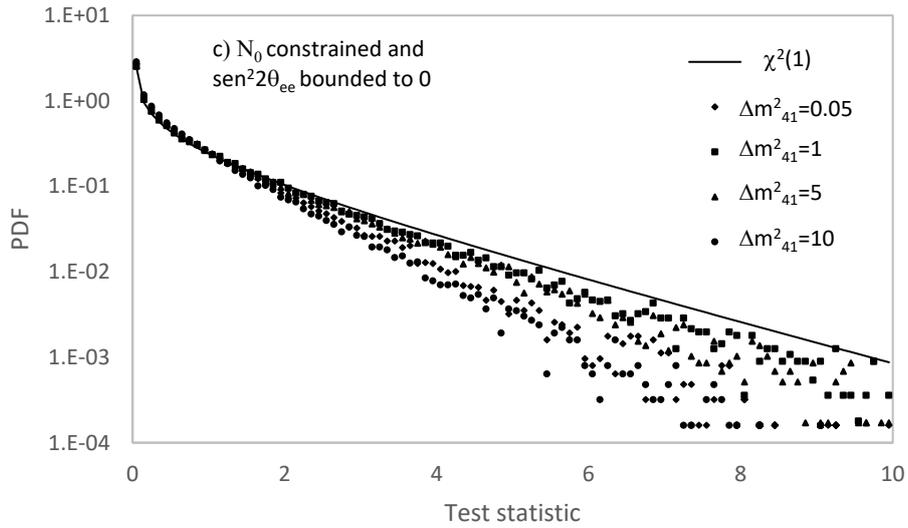

**Figure 6** - *Test statistic MC distributions for four $\Delta m^2_{41}$ values compared with the $\chi^2(1)$ function. The simulations are performed for $N_0$ constrained and $\sin^2 2\theta_{ee}$ bounded to 0. The agreement obtained in the previous two cases is now lost, though a vague resemblance with the $\chi^2(1)$ distribution persists*

Indeed, the MC distributions of the test statistic vary as function of $\Delta m^2_{41}$, do not follow any more the profile of the $\chi^2(1)$ distribution, though vaguely resemble it, and, while not shown in the figure for simplicity, exhibit a Dirac delta at zero also of varying amplitude, depending upon $\Delta m^2_{41}$ and typically smaller than the 0.5 value of the previous case.

It should be highlighted that the deviation from the $\chi^2(1)$ profile, in particular, causes a suppression of the long tail of the distribution, underlying that the effect of the application of the $N_0$ constraint is to limit the more extreme fluctuations of the test statistic, likely because of the correlation among the bins introduced by the constraint itself.

## 5. Global distribution of the test statistic

To understand the global statistical picture, let's now remove the condition of fixed $\Delta m^2_{41}$ so to allow the full manifestation of the Look Elsewhere Effect over the $\Delta m^2_{41}$ range of interest.

Before going ahead with the simulations, it is useful to recall the model to describe the distribution of a generalized likelihood ratio discovery test statistic introduced in [13] in the paradigmatic situation of the search for a particle of unknown mass over a wide range, under the null hypothesis of background only:

$$p(t_0) = N \int_0^{t_0} \chi_s^2(h)dh \left( \int_0^{t_0} \chi_{s+1}^2(h)dh \right)^{N-1} \chi_{s+1}^2(t_0) + \chi_s^2(t_0) \left( \int_0^{t_0} \chi_{s+1}^2(h)dh \right)^N \quad (6)$$

where $t_0$ represents the test statistic random variable and s is equal to 1 since locally the distribution is assumed $\chi^2(1)$ distributed, and N is obtained by comparison with the MC output.

The origin of equation (6) is a suggestive interpretation of the Look Elsewhere Effect: it actually amounts to explain the test statistic as the maximum among N+1 random variables, N distributed according to $\chi^2(2)$ and 1 as a $\chi^2(1)$. Thus, equation (6) is a so-called extreme value distribution, and heuristically it means that to find the sought after effect one looks over N independent search regions, each characterized by a test statistic $\chi^2(2)$ distributed, identifying the largest among them as alleged indication of the existence of the signal. The additional $\chi^2(1)$ variable stems from the search at the fixed border of the interval of interest.

A more detailed description of the model (6) and of its derivation can be found in [13]. It is useful to underline here that the number N of independent scanned regions cannot be predicted a-priori and can be only inferred by comparison of the MC output with the model (6) itself. Moreover, the same value N is that obtained with the alternative LEE MC prescription introduced in [12].

Finally, it must be pointed out that, as confirmed by the tests reported in [13], the model (6) is expected to be exactly valid in the case of the generalization of the unidimensional simulations realized with the conditions of paragraph 4.a, when the local distribution is actually a true $\chi^2(1)$ function, but it will be interesting to check its adherence also to the distributions obtained according to the prescriptions of paragraphs 4.b and 4.c.

*5.a Unconstrained parameters*

The results of the simulation with unconstrained $N_0$ and $\sin^2 2\theta_{ee}$ over the $\Delta m^2_{41}$ search region assumed from 0 to 10 eV$^2$, the region of interest for the sterile neutrino oscillation search, is shown in Figure 7, overlapped to the model (6).

In agreement with the expectation, the model describes perfectly the test statistic MC distribution with N=13.5, thus confirming that the basic facts behind the Look Elsewhere Effect in case of the search of a particle of unknown mass are at work also here, in the search for an oscillation effect induced by an alleged sterile neutrino, and this is not surprising since the same generalized likelihood ratio test is exploited in both cases to investigate the putative effect, being the non-applicability of the Wilks' theorem the other common feature.

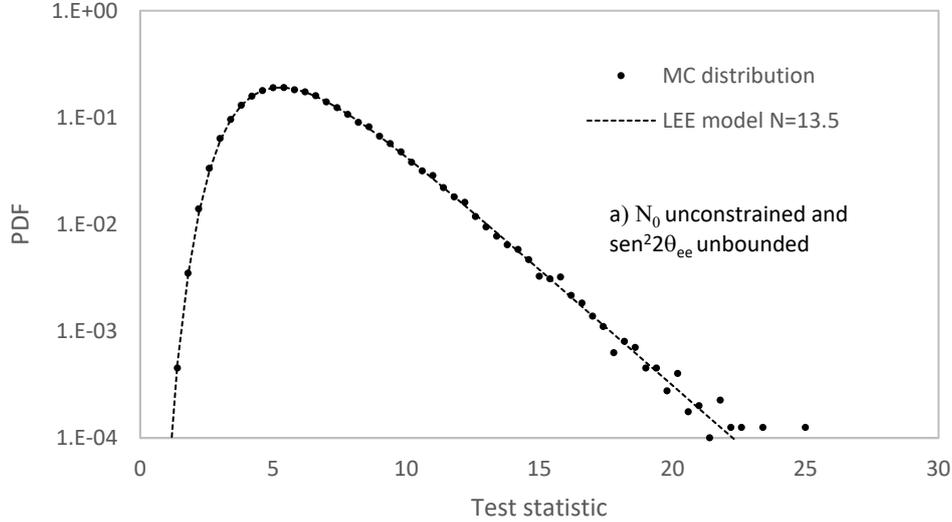

**Figure 7** – *Global test statistic MC distribution. The simulation is performed for $N_0$ unconstrained and $\sin^2 2\theta_{ee}$ unbounded, and compared with the LEE model (6) with an excellent agreement.*

In particular, the MC output, through the exact validation of model (6), provides full support to the interpretation of the test statistic as the maximum among a set of random variables with appropriate distribution.

*5.b Unconstrained $N_0$ and bounded $\sin^2 2\theta_{ee}$*

In the path towards the complete situation, the intermediate case b) with bounded $\sin^2 2\theta_{ee}$ and unconstrained $N_0$ is shown in Fig. 8. In the figure there is also for comparison the previous simulation; and the LEE model with a different N=11 to adapt to the new simulation. Two issues can be noted, that the current simulation is slightly different from the previous one, rather distorted to lower values, signaling a decrease of the "noise floor" affecting the detection of a hypothetical real signal thanks to the additional information provided through the bound on $\sin^2 2\theta_{ee}$, and that the LEE model cannot adhere exactly to the new simulation output, and this is not surprising since the local distribution is no more an exact $\chi^2(1)$ function.

However, given the small modification of the MC result, heuristically one can guess that the model can be adapted to the new configuration with some appropriate modification of the $\chi^2(2)$ building block of model (6).

To follow this possibility, it can be supposed that an appropriate modified building block can be obtained starting from a generic gamma function

$$f(x;\alpha,\beta) = \frac{1}{\beta^\alpha \Gamma(\alpha)} x^{\alpha-1} e^{-\frac{x}{\beta}} \quad (7)$$

by noting that with $\beta=2$ it reduces to a $\chi^2$ function with n degrees of freedom if $\alpha=n/2$.

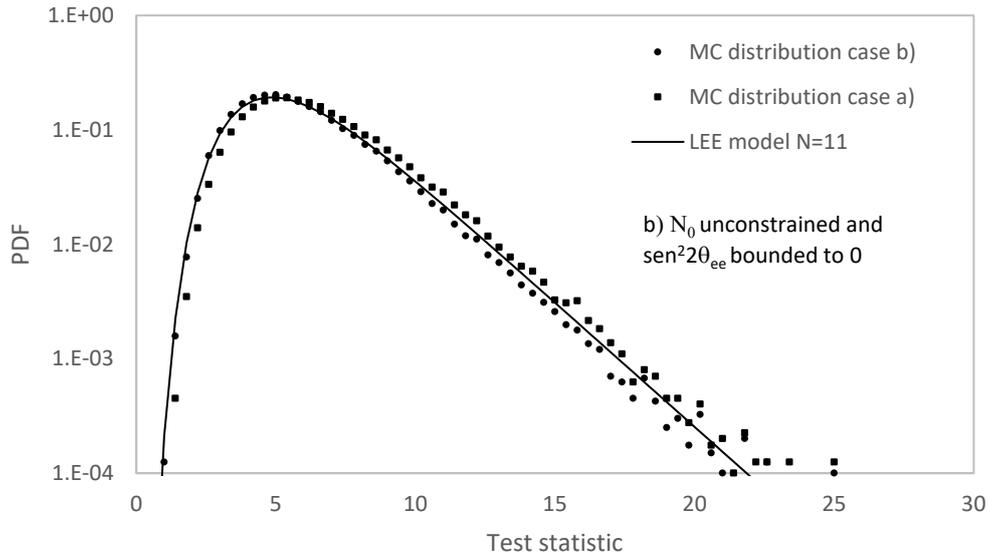

**Figure 8** – *Global test statistic MC distribution case b). The simulation is performed for $N_0$ unconstrained and $\sin^2 2\theta_{ee}$ bounded to 0, and compared with the case a) as well as with the LEE model (6), which ensures only an approximate description of the MC output.*

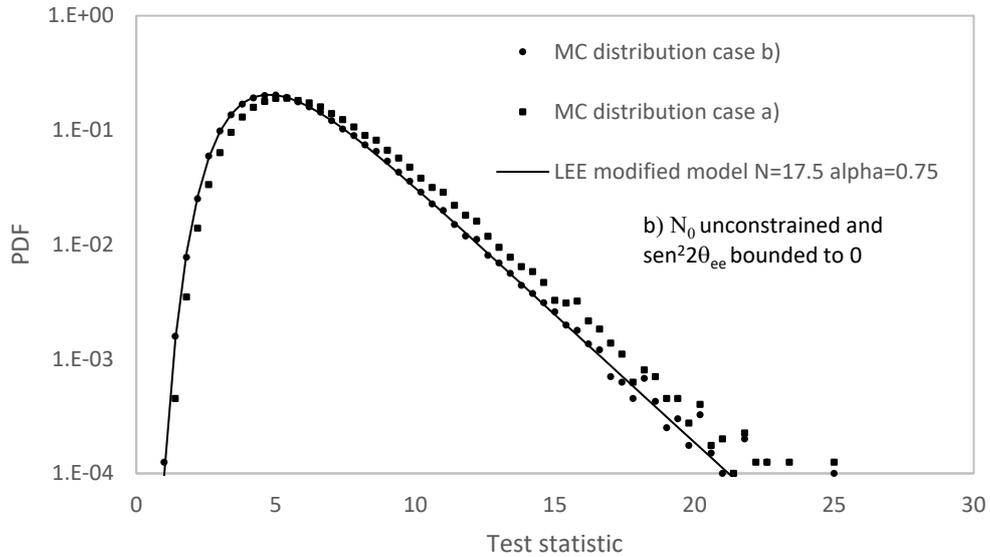

**Figure 9** - *Global test statistic MC distribution case b) compared with the modified LEE model with $N=17.5$ and $\alpha=0.75$. A rather good MC-model accord is restored.*

So, the generalization could be attempted by keeping β fixed to the value 2, and instead allowing α, which in the default model is equal to 1 to correspond to two degrees of freedom, to be free to take any value. By replacing $\chi^2(2)$ in equation (6) with the expression (7), a very good data-model match can indeed be obtained with α=0.75, as shown in figure 9. The parameter N is in this case is 17.5. Thus, basically the interpretation of the test statistic as the maximum among a certain number of random variables works rather well also in this configuration, requiring only a slight change of the distribution which they obey.

*5.c Constrained $N_0$ and constrained $sen^2 2\theta_{ee}$*

Finally, the simulation in the complete configuration is obtained by applying the constraint also to $N_0$, through the introduction of the related pull term. As counterpart of the remarkable distortion already noted while performing the simulation for fixed $\Delta m^2_{41}$, the output of the MC in this configuration is extremely distorted with respect to the two previous cases, as illustrated in figure 10.

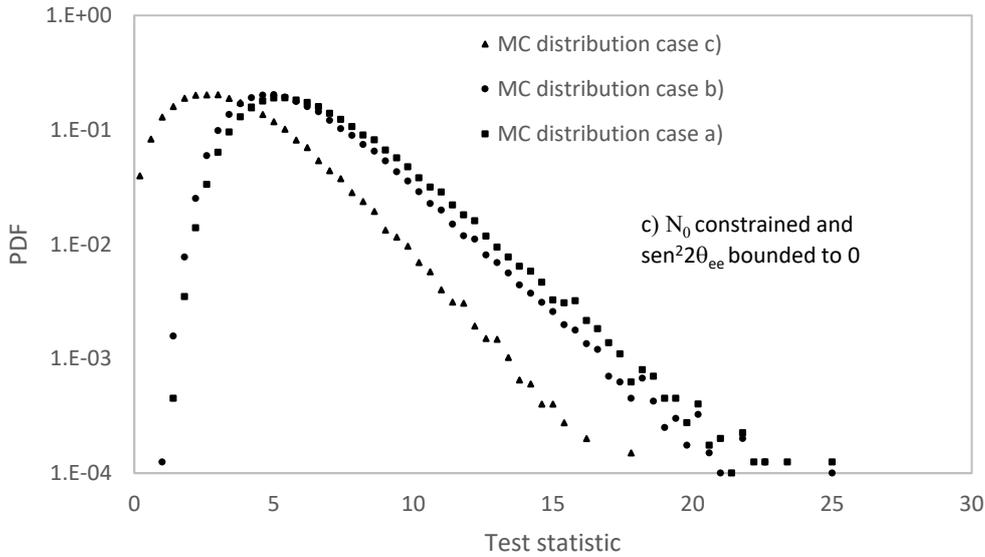

**Figure 10** - *Global test statistic MC distribution case c) compared with the previous simulations. The effect of the constraint on $N_0$ is extremely evident pushing the distribution towards the left, implying a substantial reduction in the "noise floor" for the detection of the sought after effect.*

In particular, in the complete case the tail of the test statistic is greatly suppressed, and this effect is clearly the global consequence of the damping of the tails of the local distributions, which in turn stems from the constraint on the $N_0$ nuisance parameter. Indeed, such a constraint introduces a strong correlation among the bins, substantially hindering the free fluctuations of the test statistic and somehow clipping its possible extreme values.

Overall, hence, the strong constraint on $N_0$ further reduces in a substantial way the "noise floor" affecting the detection procedure of a true effect, highlighting the beneficial consequence of the addition in the GLR formulation of external constraints reflecting any a-priori knowledge available related to the problem at hand.

Because of the severe distortion of the distribution, the model (6) cannot faithfully reproduce it. However, the bulk of the distribution can still be reasonably described, as shown in Figure 11. On the other hand, the comparison with the model further highlights the behavior of the tail, whose substantial drop with respect to the extrapolated profile underlines also visually the role of the constraint on $N_0$ in limiting the extreme values of the test statistic. Anyhow, the model itself constitutes a useful upper limit for the tail, which could be exploited to determine a lower limit on the significance in case of an extreme outcome in a real measurement, well beyond the region where the test statistic can be computed numerically through the MC calculation.

Actually, it is worth to point out in this context that the main motivation of the model (6) developed in [13], as well as of the alternative procedure illustrated in [12], was to find a way to extrapolate the tail of the test statistic well beyond the region that could be evaluated by MC, in cases contemplating significances of 5 σ and beyond, which would require an enormous amount of simulations to reasonably compute the tail at the relevant level of precision. In the disappearance sterile neutrino searches, such a situation occurs for the combined Gallium data, when including the recent results of the BEST [19] experiment a 5 σ significance effect becomes manifest [20].

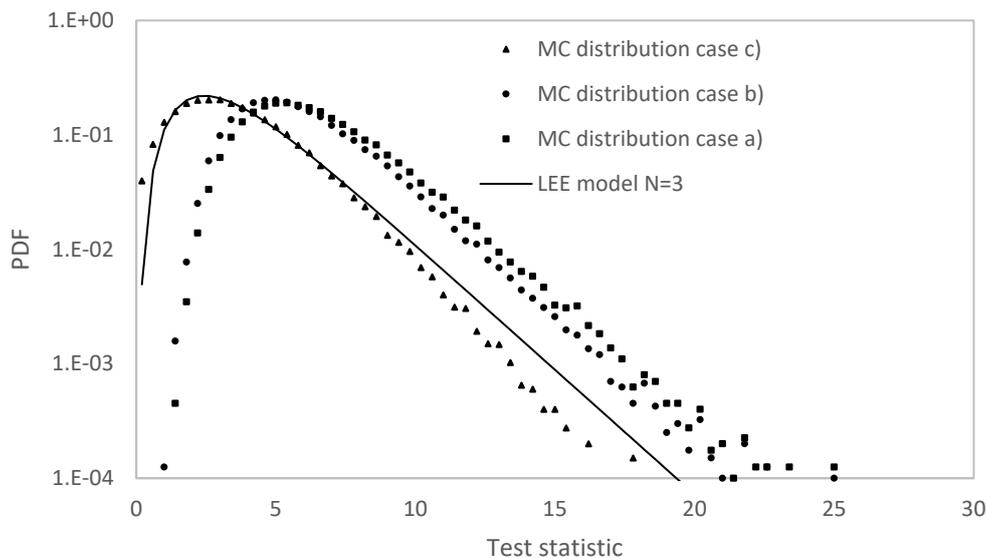

**Figure 11** – *Comparison of the global test statistic MC distribution in the complete case c) with the LEE model for N=3. Because of the strong distortion of the distribution there is only a broad agreement between the two, useful anyhow to set a lower limit on the significance of a putative detection.*

Before closing this discussion, it is appropriate to recall that the same effect on the discovery test statistic of being pushed to lower values while adding more information in the GLR, is reported also in [2], in the comparison of the test statistic computed for a paradigmatic reactor neutrino experiment either with the spectrum shape information only or also adding the extra information coming from the associated rate measurement (Fig. 8a of [2]).

## 6. Implications of the Look Elsewhere Effect in the interpretation of the results of a sterile neutrino search

The discovery test statistic (5), i.e. the highest peak in Fig. 3, in presence of only noise, that is only Poisson variation of the counts, thus follows the distributions in figure 10 depending upon the assumptions of the simulations, with particularly the triangle and circular dots curves fulfilling realistic physics conditions. Moreover, it obeys, more or less precisely, to the model (6), again depending upon the conditions. This means that without accounting for the correct distribution, the largest peak of Fig. 3 can be easily mistaken as indication of a real signal, because of some extreme noise induced fluctuation that it could feature, but not so extreme to escape the tail of relation (6) or of the related MC distributions. And the long tail of (6), as that of the MC outputs, explains, vice versa, why it is not improbable to perform an active-sterile experimental search and find some subtle "intriguing" effect, which in reality is only a manifestation of a noise induced fluctuation.

The possible confusion is exacerbated by the circumstance that naively one can be tempted to confront the maximum value of the test statistic stemming from an experiment, i.e. the largest among the peaks in a outcome like that in Fig. 3, with a $\chi^2(2)$ distribution and infer from this comparison the mathematical significance of the presumptive hinted effect, assuming that such a distribution is a faithful model of the PDF of the test statistic (the largest peak in Fig. 3) on the basis of the Wilks' theorem with two extra parameters in the second term of (5).

After the long above discussion, it is obvious that this is not the case, and such an occurrence is further confirmed by Fig. 12, where the actual distribution of the test statistic from the example test set-up, case c), is confronted with the $\chi^2(2)$ function. For completeness also, the two distributions obtained under simplified assumptions about the constraints are displayed in the same figure.

Clearly, the real and $\chi^2(2)$ distributions are rather different, and the latter underestimates the tail, thus originating a stronger significance if erroneously used to assess the probability of the highest peak of the test statistic to be larger than the actually observed value in the experimental outcome, according to the definition of p-value of a putative detection. However, the figure stresses again that in the present exercise the introduction in (5) of the pull factor to constraint the strength of the neutrino source, $N_0$, creates a strong suppression of the tail of the test statistic itself, alleviating the discrepancy with the $\chi^2(2)$, otherwise much larger in the distributions in which the pull factor is ignored.

Anyhow, this discussion elucidates the essence of the implications of the Look Elsewhere Effect, here and in any circumstance in which it appears: it consists in assuming that the discovery test statistic used as indication for the existence of a signal follows a PDF different, and with less tail, from the real one, and the latter typically is an "extreme distribution" which governs the largest value among a set of variables. On the contrary, the naively assumed PDF, if the LEE is not taken into account, stems from the improper application of the asymptotic properties stated by the Wilks' theorem and does not account for the high values that the test statistic can assume even in case of no signal.

## 7. Comparison with previous results

The above results have been obtained with a simplified, ideal experimental set-up, suited however to catch the essence of the mathematics behind the statistical ascertainment of the significance of the results in a disappearance quest of a light sterile neutrino.

It would be, nevertheless, useful and interesting to compare the present outcomes with results discussed in the literature for realistic analyses of disappearance short baseline experiments.

However, only recently in [2], [3], [4] and [5] the issue of the non-validity of the Wilks' theorem has been addressed in the context of the sterile search, and so it is not common in the experimental works to find published the MC distribution of the discovery test statistic. A paper in which such a distribution has been explicitly reported is [5], where the authors have included the discovery test statistic stemming from the joint analysis of all reactor experiments searching for a sterile neutrino.

In figure 13 this distribution, grey dots, is shown overlapped to the distributions obtained in the present exercise. The general agreement exhibited with those related to case a) and b) discussed here is remarkable, especially considering the complexity and completeness of the analysis in [5] with respect to the simple set-up adopted for the calculations presented in this work. This confirms what said at the beginning that the scheme adopted to exemplify the problem actually captures rather well the mathematics behind the statistical studies of the significance in the sterile search with disappearance experiments.

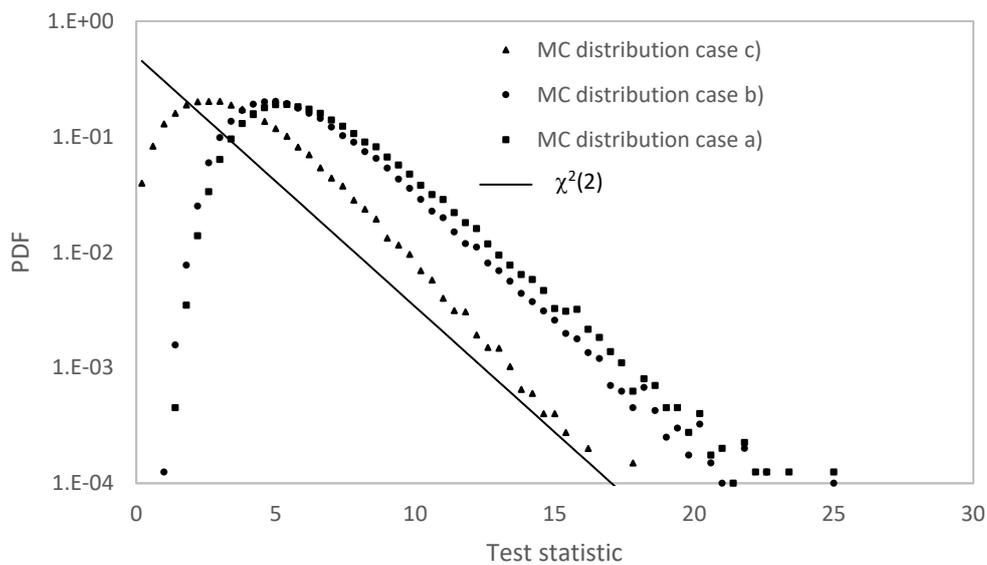

**Figure 12** – *Comparison of the global test statistic MC distributions with the $\chi^2(2)$ function. In all situations the actual tail is larger than the ideal $\chi^2(2)$ case, though the effect of the $N_0$ constraint is to alleviate such a discrepancy for the distribution obtained in the complete case c).*

A closer look shows that the all-reactors distribution follows in the first part the case b) MC simulation of this paper and in the second part the case a) MC simulation. Given these similarities, it can be useful to shed more light on the characteristics of the all-reactors distribution by verifying its accord with the LEE model, standard or modified.

In Fig. 14 the comparison with the standard LEE model is reported. Surely the comparison is rather satisfactorily with N=10.5, though the tail results slightly underestimated. On the basis of the previous observations in the context of the exemplificative set-up, this lack of perfect agreement is plausibly due to the concurrent influence of the 0 bound imposed to $sen^2 2\theta_{ee}$ and of the systematic effects expressed via the nuisance parameters included in the analysis.

Then a step forward is the application of the modified LEE model in which the $\chi^2$ function is replaced by the gamma function, with an adjusted $\alpha$ parameter in (7), reminding that in the example results the small discrepancy between case a) and b) can be compensated by this ad hoc adjustment of the model, and so it is worth to see if this is valid also for a real experimental distribution. Figure 15, actually, demonstrates that also in this case the modification of the model greatly improves the agreement with the test statistic distribution.

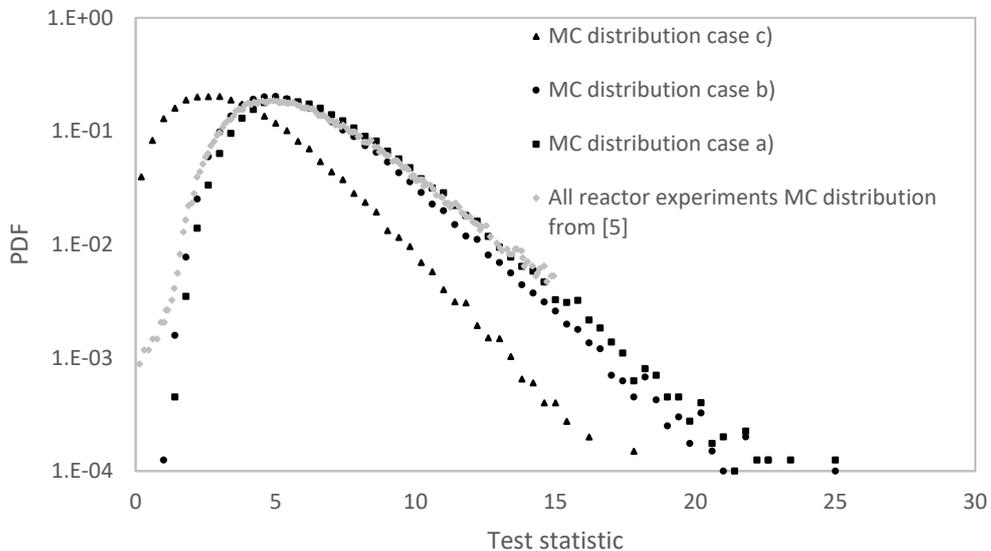

**Figure 13** – *All-reactor experiments discovery test statistic MC distribution from [5] compared with those stemming from the example set-up under the three different examined conditions. The real experimental distribution compares well with the case a) and b) of the example set-up.*

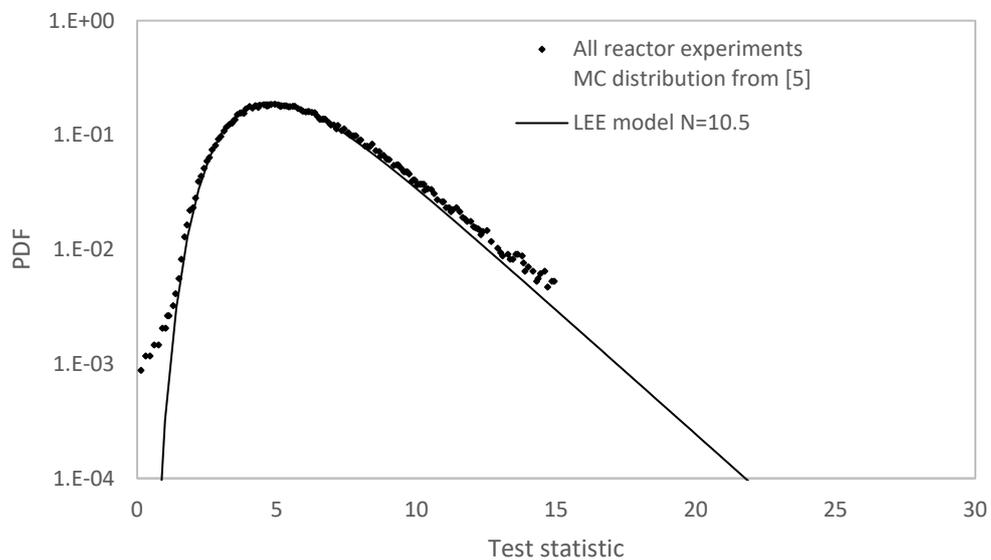

**Figure 14** – *LEE model (6) computed for the effective parameter N=10.5 to obtain the best match with the all-reactor experiments discovery test statistic distribution. The agreement is rather satisfactory, though the tail is underestimated.*

Overall these comparisons, taking into account the results of the exemplificative exercise of the previous paragraphs, allow to draw the following conclusions: 1-also the discovery test statistic of a real analysis of the disappearance reactor experiments obeys to model (6), essentially based on the maximum among a number of $\chi^2(2)$ variables; 2- in real life the shape of the distribution follows the realistic case b) of the exercise and thus the slight modification introduced of the exact LEE model is enough to precisely describe its actual profile; 3- the nuisance parameters in a real situation originate a weaker distortion of the shape of the distribution with respect to the example, and its difference with the naively assumed $\chi^2(2)$ function is therefore larger with respect to the nuisance parameter case c) of the test set up.

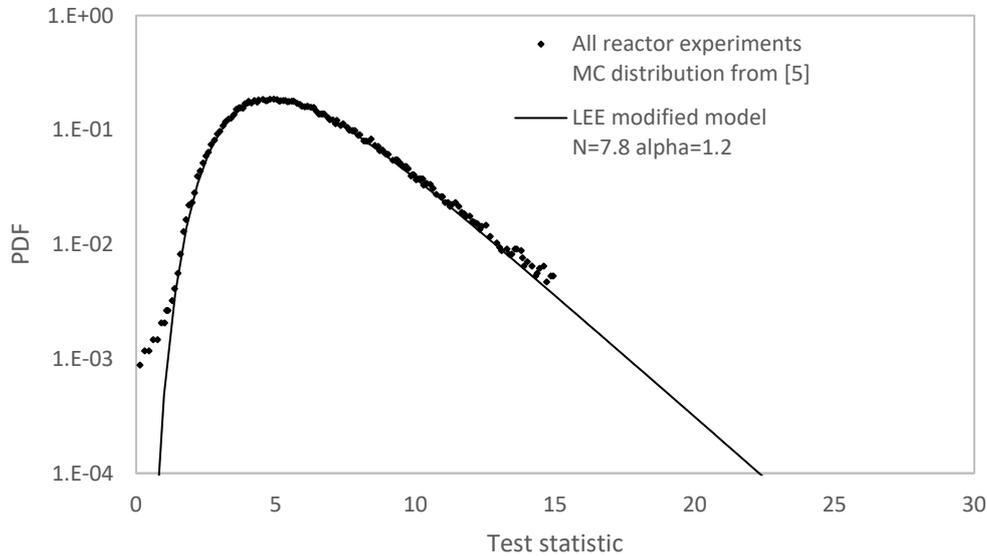

**Figure 15** – *Modified LEE model computed for the effective parameters N=7.8 and α=1.2 in (7) to improve the match with the tail of the all-reactor experiments discovery test statistic distribution. The description of the tail is definitively improved with respect to the standard LEE model in Fig. 14.*

### 8. Comparison with the square root of test statistic model

In reference [4] a model has been introduced for the distribution of the discovery test statistic in case of the active sterile oscillation search, based as well on the maximum among a number of N random variables. Specifically, in this model is the square root of the test statistic itself to be interpreted as the maximum among a number of variables, each being a standard gaussian variable.

Now, by noting that the square of a standard gaussian variable is a $\chi^2(1)$ variable, and that the peaks in figure 3 are $\chi^2(1)$ or approximately $\chi^2(1)$ distributed, such a model amounts to assume that the fluctuating variables are just these peaks. In the LEE model described above instead the fluctuating variables are $\chi^2(2)$ and are not immediately visualizable. A common feature of both models is that the number of fluctuating variables can be determined only a posteriori through the MC-model match. The main difference is that one is based on $\chi^2(1)$ building blocks and the other mainly on $\chi^2(2)$ functions.

Before proceeding with this discussion, it is useful to show concretely that modeling the root square of the test statistic using the maximum out of gaussian variables and modeling the test statistic itself with the maximum out of $\chi^2(1)$ variables is the same procedure. The perfect coincidence of the two approaches can be appreciated by plotting the test statistic distribution according to the two prescriptions and showing how it is possible to pass from one to the other.

To this purpose let's point out that given N variables with equal PDF p(t) and defining

$$F(t) = \int_0^h p(\lambda)d\lambda \qquad (8)$$

then the PDF of the highest value among them is

$$p_{highest}(t) = NF(t)^{(N-1)}p(t) \qquad (9)$$

The meaning of equation (9) is the following: by identifying one out the N variables as the largest, then the probability density of such an occurrence is its individual PDF p(t), multiplied by the previous factor $F(t)^{(N-1)}$, which expresses the probability that none of the other N-1 variables exceeds it. Since, however, such a configuration can be replicated N times, given that any of the variables can play the role of the largest, the N multiplicative factor is introduced to account for all such possible configurations.

Relation (6), expressing the LEE model adopted to describe the MC distributions derived previously, is actually a modification of the simple equation (9), written for p(t) equal to $\chi^2(2)$ and including an additional $\chi^2(1)$ variable, reflecting the effect at the border of the allowed mass region.

Equipped with eq. (9), we can now demonstrate that modeling the square root of the test statistic as the maximum of gaussian standard variables is fully equivalent to model the test statistic with the maximum of $\chi^2(1)$ variables. In Fig. 16 the PDF described by the model (9) with p(t) taken as $\chi^2(1)$ is plotted with the solid line for N=60. With a standard variable transformation procedure, accounting for the relevant Jacobian, from the solid curve it is obtained the dashed PDF of the square root of the test statistic. On the other hand, the same PDF of the square root of the test statistic, black dots, is obtained by calculating directly the distribution of the maximum among a set of standard gaussian variables. This is done through the same model (9), but with p(t) now taken as a standard gaussian function, and with N=120. The overlap is extremely good, concretely showing the perfect agreement of the two models.

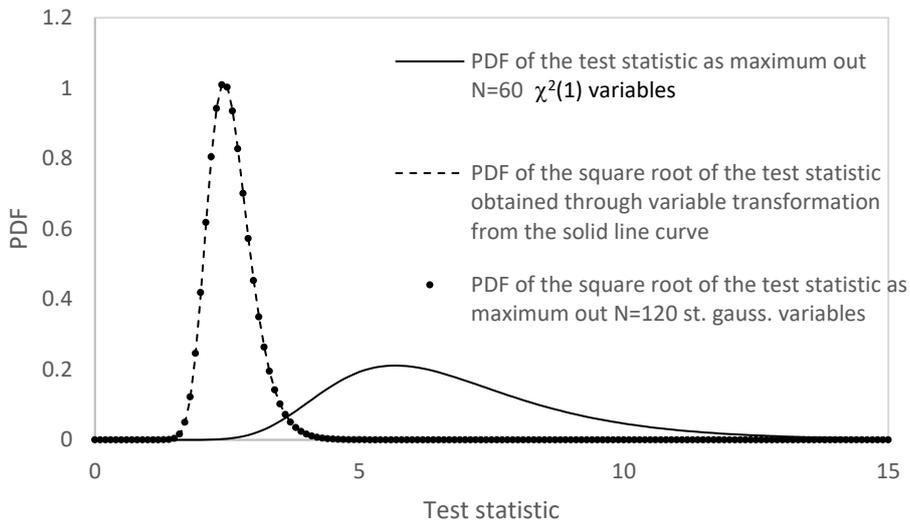

**Figure 16** – *The modeling of the square root of the test statistic as maximum out N standard gaussian variables is equivalent to model the test statistic itself with the maximum out of N/2 $\chi^2(1)$ variables, as shown by the perfect overlap of the same square root test statistic distribution obtained directly (dots) or indirectly through the $\chi^2(1)$ model (dashed curve).*

The reason why the maximum over a set of gaussian standard variables for the description of the square root of the test statistic requires a value of N double of that of the corresponding $\chi^2(1)$ model of the test statistic itself is because, when considering the extreme high values among a set of standard gaussian variables, only the right, positive part of the gaussian function matters, while the maximum among $\chi^2(1)$ variables, since they come from the square of the gaussian functions, implicitly implies extreme values from both side of the gaussian distribution, and this explains the factor 2 difference in the respective N parameters.

We can now compare the behavior of the two models by confronting them with a same MC distribution of the test statistic. It could be interesting to do it first for the most ideal distribution stemming from the exemplificative exercise, i.e. the test statistic studied in the case 5a).

The comparison is shown in Fig. 17 and clearly the $\chi^2(1)$ model can broadly describe the features of the simulated distribution, and in this sense is surely a useful approximation, but underperforms with respect to the optimal match of the LEE model (6). In particular, the first part of the simulation curve can be described by the $\chi^2(1)$ model with a lower value of the number N of variables involved in the search of the maximum, i.e. 40, while the tail requires the much greater value of 70.

Another interesting comparison is that with the all-reactor experiments distribution from [5] reported in figure 13. This comparison is shown in Fig. 18, for the two values of N of 40 and 60, which have been selected as counterpart of the modeling of the square root of the test statistic performed in [5] with the maximum out standard gaussian variables, carried out with N respectively equal to 80 and 120. Remind, indeed, that the mapping to the $\chi^2(1)$ modeling implies to adopt half of these values.

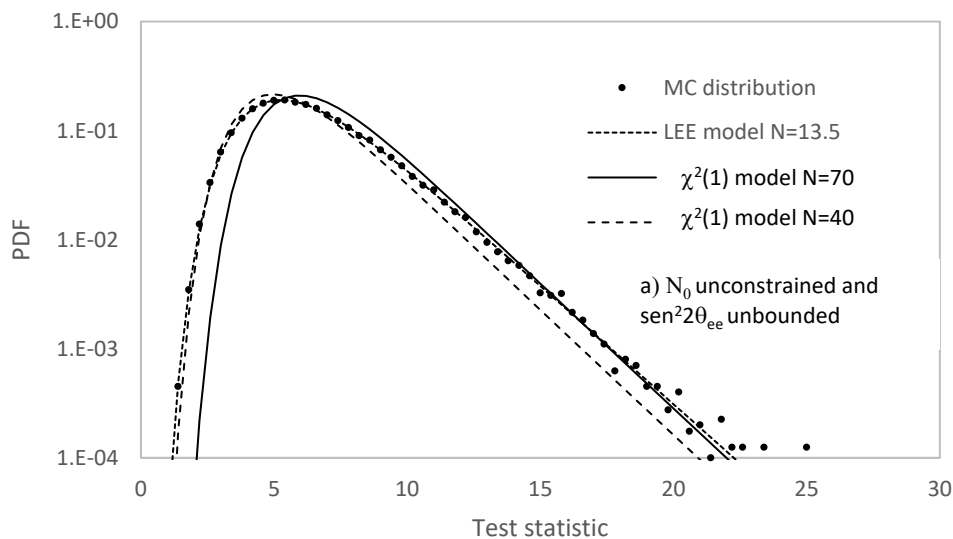

**Figure 17** – *The MC distribution of the test statistic of the example test set-up case a) overlapped to the LEE model and compared with the $\chi^2(1)$ model for the two N values of 70, suited for the tail, and 40, better describing the bulk of the distribution.*

Remarkably, the situation is rather similar to that in Fig. 17: also for this realistic distribution the maximum of $\chi^2(1)$ variables reproduces broadly its features, N=40 better adhering to the bulk and N=60 more adequate for the tail, while the LEE model (for simplicity we show only that slightly modified of fig. 15) instead exhibits a much closer overall accord.

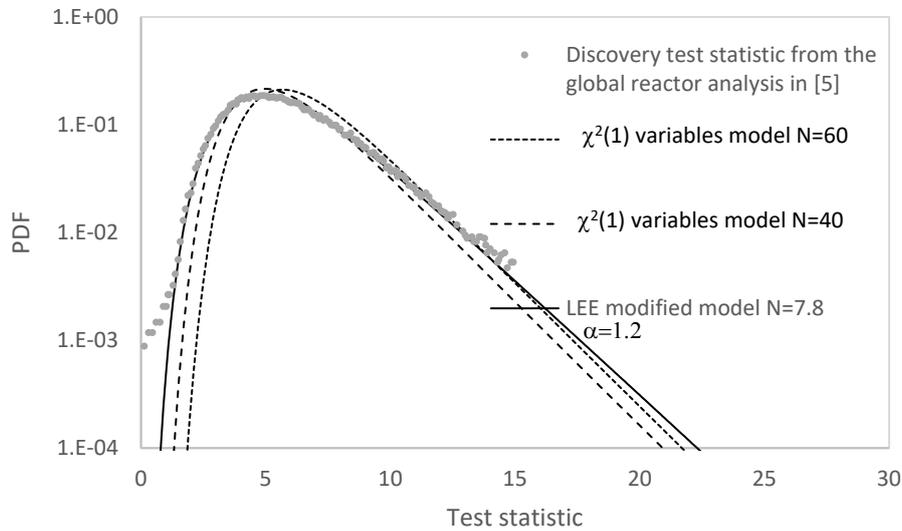

**Figure 18** - *The all-reactor experiments test statistic MC distribution from [5] overlapped to the modified LEE model and compared with the $\chi^2(1)$ model for the two N values of 40 and 60, inferred from [5].*

Finally, it can be shown how to exploit the LEE model for the motivation for which it has been originally developed in [13], i.e. the extrapolation to significance values unattainable by MC calculation, resorting to the outcomes of the reactors+Gallium study shown in the same reference [5].

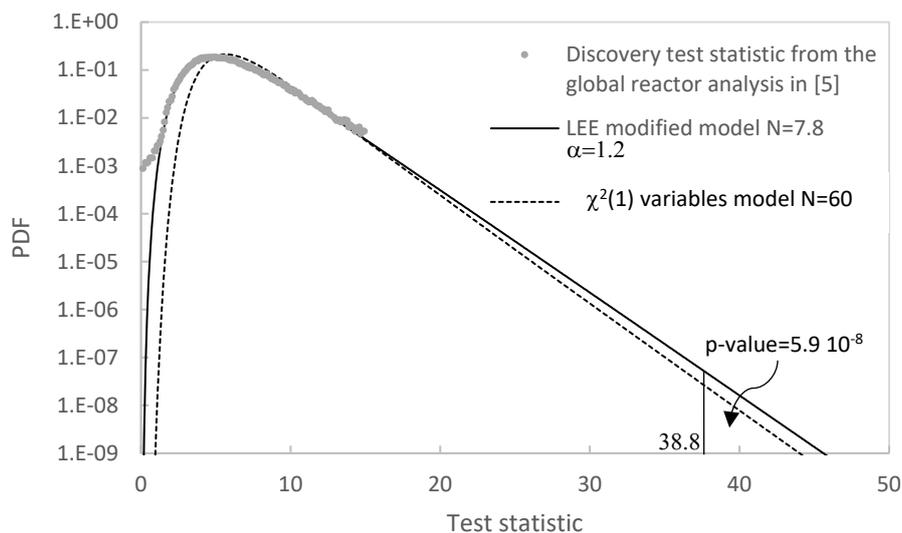

**Figure 19** – *Extrapolation of the LEE model and of the c2(1) model to large values of the test statistic and comparison with the experimental outcome of the joint reactors+Gallium analysis of [5].*

The authors of [5] find that the test statistic value stemming from the reactors+Gallium joint analysis is 38.8 and that the relevant MC distribution is essentially unaltered with respect to the reactors only case. The LEE model reported in Fig. 18 can thus be extrapolated to such a large value of the test statistic, as illustrated in Fig. 19, and the p-value of the detection calculated as the integral of the tail above 38.8, amounting to $5.9 \cdot 10^{-8}$, which converted to number of $\sigma$, with a two tailed convention, corresponds to $5.4\,\sigma$. By the way, such a large significance is the effect of the BEST results [19].

In [5] the extrapolation is instead attempted with the square root of the test statistic model, and the same calculation is repeated here with the equivalent $\chi^2(1)$ model with N=60, also plotted in Fig. 19, obtaining above 38.8 the p-value of $2.8 \cdot 10^{-8}$, corresponding to $5.6\,\sigma$, close to $5.7\,\sigma$ obtained in 5.

The LEE model, able to precisely describe the bulk of the MC distribution, allows thus an unambiguous extrapolation to large significances, but however also the approximated $\chi^2(1)$ model provides an acceptable way to extrapolate to great values of the test statistic, provided that the model is properly calibrated to the trailing edge of the MC distribution.

## 9. Conclusions

The main outcome of this work is that the determination of the significance in the light sterile search with disappearance experiments follows the same paradigm of the Look Elsewhere Effect LEE now well established in other fields of HEP. It has been shown that its manifestation also in the present context stems from intrinsic features of the generalized likelihood ratio test (also referred as profile likelihood ratio test) used to address the statistical processing of the data coming from a disappearance search as a specific problem of composite hypotheses testing, and that in particular is linked to the non-validity of the Wilks' theorem when one of the parameters, $\Delta m^2_{41}$ in this case, exists only under the alternative.

Mathematically, the LEE leads to replace the simple $\chi^2(2)$ form stemming from the Wilks' theorem under the null hypothesis of background only, with a more complex functional expression of the test statistic distribution, the model (6), coming from its interpretation as the maximum out of a certain number of random variables individually $\chi^2(2)$ distributed, plus one of $\chi^2(1)$ form. While the predictive power of this model does not allow to determine a priori the number of effective fluctuating variables, which have to be evaluated via MC, however the resulting shape, featuring a tail well in excess of the $\chi^2(2)$ function, explains why it is not uncommon to perform a sterile disappearance search and find some intriguing indication for an apparent oscillation effect, instead due to a noise induced fluctuation. Vice versa, the proper exploitation of the LEE test statistic allows to correctly assess the significance of a putative detection, avoiding the overstatement stemming from the improper use of the non-relevant $\chi^2(2)$ distribution.

The practical implications of the LEE have been thoroughly illustrated with a typical example of a disappearance set-up and then applied to a concrete analysis presented in the literature, showing that the developed framework reproduces quite accurately the features of the test statistic distribution inferred from a complete reactor experiments analysis.

**Acknowledgements**

I would like to thank Aldo Ianni and Marco Pallavicini for the development of the mathematical framework of a source-based test, which has been adapted to the example set-up used for the statistical calculations reported in this work, Alessandra Re for assistance with the computing resources, and Nicola Rossi for his help in the production of the figures.